**A machine learning framework for computationally expensive transient models**

Prashant Kumar, Kushal Sinha*, Nandkishor Nere, Yujin Shin, Raimundo Ho, Ahmad Y. Sheikh, Laurie Mlinar

[1]Process Research and Development, AbbVie Inc., North Chicago, Illinois, USA

*Corresponding Author: kushal.sinha@abbvie.com

**Abstract**

The promise of machine learning has been explored in a variety of scientific disciplines in the last few years, however, its application on first-principles based computationally expensive tools is still in nascent stage. Even with the advances in computational resources and power, transient simulations of large-scale dynamic systems using a variety of the first-principles based computational tools are still limited. In this work, we propose an ensemble approach where we combine one such computationally expensive tool, called discrete element method (DEM), with a time-series forecasting method called auto-regressive integrated moving average (ARIMA) and machine-learning methods to significantly reduce the computational burden while retaining model accuracy and performance. The developed machine-learning model shows good predictability and agreement with the literature, demonstrating its tremendous potential in scientific computing.

**Introduction**

Machine learning has emerged as one of the most promising technologies in the past decade due to its capability to provide valuable insights into vast amounts of data generated during the Internet era. Rapid democratization of machine learning tools has allowed for the successful adoption of the technology in a wide range of fields including robotics, computer vision[1], speech and natural language processing[2], autonomous driving[3], neuroscience, drug-discovery[4] and in fundamental sciences[5]. However, its application to the computational sciences, and applied computational physics in general, has been limited. Prior efforts to apply machine learning to computational sciences have primarily focused on steady state problems which are more tractable. However, applications of machine learning to time-variant problems are rare.

Over the past decade, a tremendous growth in computational power, easily accessed through cloud computing platforms, has been observed. Even then, simulations based on first-principles models of natural systems and in particular, time-variant problems of these systems remain prohibitively expensive for most practical problems. Many of these models such as Molecular Dynamics (MD)[6] used for enhancing understanding of molecular arrangements, Computational Fluid Dynamics (CFD)[7] used for understanding flow patterns for both gas and liquid phase, Density Functional Theory (DFT)[8] used for understanding electronic structure, Discrete Element Methods (DEM)[9] used for understanding particulate systems and last but not the least Finite Element Method (FEM)[10] used to measure structural strength of materials, have immense potential to accelerate research in fields impacted by them and ultimately change the world around us. Advances in the field of ML and deep learning combined with its rapid democratization, adoption in other fields and ultimately fueled by the rapid growth of computational power in the form of on-demand cloud computing certainly creates an opportunity for ML to be utilized for high-fidelity scientific computing as shown in Fig 1. Described framework allows for the development of more accurate system maps using ML tools which can be utilized for optimization and decision-making.

Within all of the fields outlined, due to extremely high computational requirements in terms of time and resources, in addition to several simplifying assumptions, one has to rely on more affordable coarse-grained systems to characterize or predict the overall state of a complex system or its relevant properties which ultimately may limit the accuracy of the results. Another way of enhancing our understanding with these high-fidelity models is to apply them under idealized conditions or in some regimes of interest. Even then time-variant simulations of these simplified models are highly expensive as well as unstable, due to the restrictions on time steps and other process parameters, to be used for most practical problems. The temporal component of many of these relevant problems are either neglected in high-fidelity models or are solved through highly idealized systems of ordinary differential equations which may ignore a lot of relevant details. Most practically relevant transient problems require simulations on the order of hours or days, however stability and computational burden only allows for a few minutes of simulations. Zonal or multi-compartmental modeling[11,12] has been used in some areas such as CFD simulations to overcome the computational cost of transient simulations, however the inherent problem with all these approaches is the difficulty in defining the zones that efficiently capture the flow behavior. A great deal of opportunity exists if we can efficiently learn the behavior of the system from a few time-steps (completed in a feasible computational time) and forecast it in time space to remove the need for running these computationally expensive simulations until completion.

In light of the above-mentioned challenges, there is an obvious need for a broadly applicable ensemble modeling framework to overcome computational limitations and move towards high fidelity predictive models that can bridge the gap between coarse-grained systems and real systems in a computationally affordable manner. In order to overcome the challenges of performing transient simulations, we propose the use of a time-series forecasting method known as auto-regressive integrated moving average (ARIMA). ARIMA has been previously used in weather forecasting and stock market prediction; however, its application to first principle models has not been reported so far to the best of our knowledge. ARIMA can be used to train on data generated from high fidelity transient simulations and then forecast key relevant physical quantities of interest. As ARIMA is learning from the entire simulation dataset, it has capability to capture start-up transients, local heterogeneities and temporal evolution of the solution. ARIMA can be an excellent tool to probe when the real system under investigation will reach to the desired state, and also the spatial distribution of time-variant physical quantities at that desired state. Taking it a step further, a machine learning predictive model can be built on ARIMA results for reaching the desired state as a function of multidimensional system parameters. Machine learning models are quick to probe and preserve the information of high fidelity models making it an excellent tool for real-time analysis, optimization, and model-based control of the system of interest.

We selected particulate mixing as our test problem for framework development due to its broad applicability in pharmaceutical, food and agro-sciences industries. Solid particles mixing is indispensable to achieve desired product quality with respect to content uniformity and reproducible manufacturing across scales in many industrial processes such as drying, blending and granulation [13,14]. Across the aforementioned applications, understanding of mixing also renders optimal process design and robust scale-up. Controlled mixing can reduce the process cycle time by multiple folds and decrease hazards such as particle agglomeration or breakage due to attrition to ensure optimum product quality. Solid particulate matter and associated processes are complex due to the factors such as single particle properties, equipment design, and modes of mixing[14].

We will further focus our attention to applications of particulate mixing in the pharmaceutical industry. Downstream process development of pharmaceutical solids, also known as the active pharmaceutical ingredient or API, involve a variety of complex chemical and physical unit operations, some of which are poorly understood. Post reaction and crystallization steps, slurry is pressure transferred to an agitated filter dryer (AFD) from the crystallizer for filtration and drying as the active pharmaceutical ingredient (API) obtained during the crystallization step is typically potent and thus human-exposure should be minimized. For drying, heat is provided from jacketed vessel walls. Intermittent agitation is usually performed to achieve uniform heat transfer across the API bed. An unoptimized agitation protocol will lead to potential agglomeration and/or attrition which would significantly impact the particle size distribution (PSD) achieved at the end of the drying and required for oral drug product formulation downstream. Particulate mixing also impacts the blending and granulation steps right before tableting and thus impact the content uniformity and drugability of each tablet. Unfortunately, though very critical, particulate mixing is a poorly understood phenomena.

A first-principles modeling technique such as DEM can reveal the underlying mechanistic aspect of particulate mixing, however like all other high-fidelity scientific computing techniques discussed above, DEM also suffers from the requirement of enormous computing power as practical systems of interest are quite large. For example, API particles of 10 µm size at the manufacturing scale filter dryer (0.88 m diameter and fill level of 20 cm) yields a system comprising of more than 20 trillion particles which would take around 7000 CPU core-years to simulate one minute of physical time for particle mixing. Hence, DEM simulations are feasible and limited to systems with small number of particles or equivalently larger particles for the same fill level. Computational requirements significantly increase further if we are dealing with cohesive particles or need to simulate for longer transient periods. Thus, over the years, a large body of DEM simulations[16,17,18] performed to understand particulate mixing have limited their investigation to smaller systems.

In this work, we present an ingenious framework for utilizing ARIMA and ML models for computationally expensive transient models (Figure 2). It should be noted that a very similar route can be taken for other cases. Spatially-averaged segregation index was used in this work, however a logical extension would be to divide the domain in multiple relevant zones and track desired physical quantities as a function of time in each of these zones, perform ARIMA to predict the time required to reach a desired state and subsequently use ML to map out the entire spatiotemporal evolution of the system.

**Results**

**Segregation Index from DEM simulations**

DEM simulations of cohesive granular pharmaceutical particles were performed in a manufacturing scale agitated filter dryer equipment. DEM equations are explained in detail in supplementary section S.1. Similar systems can be found in food, agriculture, mining and chemical industries where particle or powder handling is quite common. In DEM simulations, particle motion is described in a Lagrangian framework wherein, equations of motion are solved for each particle or each particle acts as a computational node. At each time step, the forces acting on a particle are computed. A multitude of forces can be acting at the granular particle scale such as friction, contact plasticity, cohesion, adhesion, liquid bridging, gravity and electrostatics depending upon the system under study[19,20,21].

In total, 65 simulations were performed for one minute of physical agitation time by varying the material properties encompassing a range of particle radius R, particle density ρ, coefficient of restitution *e*, cohesive energy density $\gamma_{cohesion}$, tangential friction $\mu_f$, Young's modulus E and process parameters covering a span of number of particles $N_P$, impeller speed RPM, and cake height *h*. The typical average time for each of these DEM simulations was over a month. Figure 3 shows the violin plot[22] of the range and frequency of a given parameter in our simulation design space. DEM simulations are initiated with two distinct vertical layers of the particles of types 1 and 2, and the position and velocity of the particles is tracked at all times as shown in Figure 4(a).

The extent of particle mixing is quantified using the Segregation Index parameter, ψ[23], defined in Eq. 1. '$C_{ij}$' represents the total number of contacts between particles of type '*i*' and '*j*' in a given domain. ψ is equal to 1 for a uniform random mixing, whereas it is equal to 2 for a completely unmixed scenario as can be seen in Figure 4 (b).

$$\psi = \frac{C_{11}}{C_{11}+C_{12}} + \frac{C_{22}}{C_{22}+C_{21}}$$  Eq. 1

The asymptotic value of ψ for a system would tend to 1 when approaching random uniform mixing, however the time required would depend on a number of factors, which have been investigated in this study. During mixing, spatial arrangement of the particles changes with time resulting in the evolution of the segregation index. At any time, the extent of mixing of particles will be different in different regions of the domain indicating a spatial distribution of ψ as shown in Figure 4 (c). It can be attributed to the increase in the linear velocity of the particles along the radial direction resulting in differences in particle collision frequency and efficiency. Whereas in Figure 4 (d), it can be seen that bulk averaged ψ decreases with time during mixing for different impeller angular velocity. With more impeller revolutions, the bed becomes better mixed resulting in a drop in the bulk averaged value of ψ. It is clear that longer mixing times would be required at lower RPMs as mixing is driven by the number of revolutions. Even though the bulk averaged ψ approaches to one, there may be regions closer to the impeller's axis of rotation (regions R1 and R2 in Figure 4 (c)) where more agitation would be required for uniform mixing. The caveat with a longer agitation period is that it can affect particle size distribution because of particle attrition[24] and agglomeration[21]. Particle agglomeration and attrition are the key challenges that govern decisions or design of an optimized agitation protocol and need to be overcome to have a robust process. It therefore becomes crucial to know the approximate agitation time required for uniform mixing.

**Time Forecasting of Segregation Index**

Instead of simulating for the entire physical operation time which is prohibitively large, we chose to simulate for one minute of operation and project the results (segregation index with time) in time-space using a time-series forecasting method, ARIMA to overcome the prohibitively large simulation time of a high-fidelity simulation technique like DEM.

ARIMA[25] is one of the most widely used approaches for time-series forecasting in finance[26] and econometrics as it aims to describe the autocorrelation in the data for forecasting. ARIMA models can handle both seasonal and non-seasonal data and offer advantage over classical exponential smoothing methods. Spline-fitting was also implemented, but it did not perform well due to the noisy nature of the data in certain cases. Time-series data can sometimes be extremely noisy making it difficult to untangle the mean 'stationary' behavior from the noise. ARIMA can transform time-series data into 'stationary' post-differencing, or in other words, a combination of a signal and noise. The elements constituting ARIMA are the number of autoregressive terms required for good forecasting (p), the number of differencing operations to achieve stationarity (d), and the number of lagged forecast errors (q). ARIMA formulation is explained in detail in supplementary section S.2. Differencing and regression using the 'relevant' previous time points, unlike other methods, helped ARIMA to capture the non-seasonal and non-stationary behavior of the segregation index at higher RPMs.

We chose to do time-forecasting of Segregation index, $\psi$, which is an indicator of the extent of particulate mixing. ARIMA predictions were verified on all DEM generated data by training on $\psi_{t=0}$ to $\psi_{t=T/2}$, where T is the total time step of the DEM simulation and predicting on the latter half (t=T/2+1 to t=T), as can be seen in Figure 5 (a). The ARIMA model was able to capture the temporal evolution of the segregation index with an error margin of less than 2.5% from the prediction of DEM simulations. ARIMA validation is summarized in supplementary section S.4.

Post-verification, the ARIMA model was used to forecast the trend of ψ and the time required to reach the desired state of uniform random mixing, i.e. ψ ~ 1 as shown in Figure 5 (b). In this work, a cut-off of ψ =1.1 was chosen to determine uniform mixing as the asymptotically slow approach of ψ towards 1 would result in erroneously large predicted mixing time. It should be noted that the ARIMA model took computational time of *O*(minutes) while DEM simulation would have typically taken another one and half months running on the same computational resources for the result shown in Figure 5.

ARIMA, though applied here on ψ, could have been applied on another time-varying physical quantity of interest such as torque, stress, and kinetic energy depending on the needs of the study. ARIMA is a powerful tool to reduce the computational cost and time by several orders of magnitude, in terms of core-hours, as indicated in Table 1. End-point estimation using the combination of DEM and ARIMA frees-up computational resources that can now be utilized for a parameter sweep of the entire relevant range of material property and process parameters to build a robust machine learning model.

**Table 1: Computational Time of DEM and ARIMA simulations.** Computational time for DEM significantly increases with the number of particles, whereas the computational time for ARIMA is only affected by the number of previous time steps to analyze and the number of future time steps to forecast. Hence, the computational time to run ARIMA for each of the cases was on *O*(minutes).

| Particle Size, R (μm) | Number of Particles, Np | Fill Level, h (cm) | Young's Modulus (N/m²) | DEM Simulation Time (CPU hours) |
|---|---|---|---|---|

| 1500 | 250,000   | 0.69  | 1e+7   | 1584  |
| 1500 | 1,000,000 | 2.79  | 2.5e+7 | 4,248 |
| 4500 | 135,000   | 10.17 | 5e+7   | 1,685 |
| 4500 | 270,000   | 20.33 | 5e+6   | 2,160 |

**Machine Learning Predictive Model**

The desire to develop a ML model stemmed from our vision to utilize data from high-fidelity simulations for process optimization and online control. We envision a manufacturing platform where advanced process analytical tools (PAT) are feeding process data to a controller which utilizes high-fidelity simulations guided ML models for making process decisions. Once connected with PAT devices, these ML models can improve their prediction over time as more process data becomes available. ML allows learning from a large number of process descriptors along with advanced feature engineering, which enables robust predictions of systems with complex phenomenon, and also has been shown to work better than linear regression techniques [28,5,29]. Another advantage of a ML model is that it eliminates the need for running costly high-fidelity simulations in the future and provides deeper insights and patterns which were earlier non-decipherable. Although, machine learning methods are great at predicting interpolated results, they may not perform well when the values for the descriptors are far from the training set. To overcome this limitation, we created a diverse descriptor design space.

In this work, ARIMA forecasted uniform mixing time was taken as the response variable to be predicted as a function of a set of input parameters such as material and process properties.

Implementing sophisticated machine learning methods such as neural networks was tempting but not practical because of the dimensions of the dataset making it vulnerable to over-fitting. Random forest outperformed ($R^2$=0.79) the other methods because of averaging the results from multiple trees generated from a randomly selected subset of the data. Partial Least Squares Regression (PLSR), Support Vector Regression, and regularized linear regression technique Elastic Net were inferior in performance as compared to random forest with an $R^2$ of 0.70, 0.72 and 0.71 respectively, also can be seen in Figure 6. However, all these methods performed better than the conventional linear regression because of the non-linear interactions arising from the complex interplay of the underlying multi-physics phenomena. Leave-one-out cross-validation was performed on all the above investigated ML methods to test their prediction and also vulnerability to over-fitting. Further, robustness in performance can be ensured as more and more process data becomes available for integration into the existing ML models.

Having obtained the predictive ML model, we sought to gain mechanistic insight in the system by probing which descriptors impact the response variable the most by a process known as feature selection. RT-RF, our best ML model, identifies the importance of the descriptors rather than weights attached to the descriptors like a linear regression model. Importance of a feature is quantified by calculating the percentage change in mean-squared error by changing the value of the descriptor. According to RT-RF, fill level, impeller rotation and particle radius, in decreasing order of significance, are the most informative descriptors to impact uniform mixing time which is in good agreement with some recent works[30,31].

At larger fill levels, particles need to be displaced to a greater extent to achieve uniform mixing leading to an increase in mixing time[30]. In a similar manner, at the same fill level, increasing impeller speed creates larger convective diffusion and thus reduces mixing time [30]. Local shear diffusion rate scales as ~ $\dot{\gamma}a^2$, where a is the particle radius, which was also identified as a critical parameter by random forest. We hypothesize that in our system convective and shear diffusion play an important role in mixing based on these results. Though similar conclusions could have been arrived by other means, ML allows us to provide relative weight to each descriptor of the system and thus provides framework for mechanistic exploration. In a convoluted system, like the one studied here, where there are multiple descriptors and fundamental understanding is missing, ML can be a powerful tool to point theorists in the right direction.

**Discussion**

i. A large amount of resources and time are spent in a variety of industries dealing with solid handling in developing a robust, scalable and reproducible process combined with technology transfer to manufacturing sites, a lot of which happens in an ad-hoc manner. Fundamental scientific tools, though accurate, have prohibitively large computational cost, particularly for transient cases while most industrial processes, either batch or continuous, have a transient component in their operation. The study presented here shows that for a complex and relevant case of cohesive powder mixing, a novel approach based on time-series forecasting using ARIMA and ML can provide tremendous insights and guide mechanistic framework by pointing key descriptors that impact the outcome the most. The overall framework presented here is quite simple and powerful and can be adopted in a variety of engineering and scientific problems that are transient in nature. An easier extension of the framework can be done in the field of computational fluid dynamics (CFD) to probe various heat and mass transfer limited or phase transition systems. Similarly, it can be used in the field of molecular dynamics (MD) to predict the rate or final packing form of materials or biological entities like DNA or proteins. Coupling of fundamental tools with ARIMA and ML would reduce the computational time to probe a large descriptor set and provide predictions on the behavior of the system under new conditions and/or, additionally, the optimal way of operating the system. From an industrial perspective, the ML models can become part of model-predictive control and coupled with PAT and automation providing endless opportunities.

**Method Description**

**DEM Simulation**

SmartDEM (Tridiagonal Solutions, San Antonio, TX) software was employed to perform all DEM simulations. SmartDEM is a GUI implementation of the open-source DEM code LIGGGHTS (LAMMPS (Large-scale Atomic/Molecular Massively Parallel Simulator) Improved for General Granular and Granular Heat Transfer Simulations; CFDEM Project) and allows for ease of simulation setup and result interpretation. Multiple automated scripts were written to create different simulation setup, submit jobs to a super-computing cluster and post-process the gigabytes of simulated data. The details of the DEM formulation are in supplementary section S.1.

**ARIMA Forecasting**

A python code[32] was customized to forecast the segregation index. As the mixing time is a complex function of the descriptors, one ARIMA model would not work best for all the data. So, ARIMA hyperparameters (p,d,q) were sampled between 0 and 100, 0 and 2, 0 and 2 respectively for all the simulations. Given that errors at previous time steps are unobserved variables, maximum likelihood estimation (MLE) was performed in order to find the best model. Akaike Information Criterion[27] (AIC) score was used to select the best ARIMA model after comparing each model against other models. ARIMA python codes were run on the Anaconda[33] platform using jupyter notebooks. All the simulations took $O$(minutes) for completion, which reflects on the power and scalability of the method. The time complexity of ARIMA is a function of the number of values of hyperparameters to sample rather than the number of particles or the values of the other descriptors, as compared to the DEM simulations where computational time significantly depends on the fill level and the number of particles.

**Machine Learning Methods**

Machine learning methods such as Elastic Net Regression (EN)[34], Support Vector Regression (SVR)[35], Partial Least Squares Regression (PLSR)[36], and Regression Tree Random Forest (RT-RF)[37] were used to build the predictive model (refer to Figure 2). Because of the limited number of datasets available, artificial neural networks was not implemented because of the concerns of over-fitting. A variety of linear, regularized linear, and non-linear methods were evaluated, of which random forest performed the best. Leave-one-out cross-validation was performed to evaluate the machine learning methods and test the vulnerability of the methods towards over-fitting. Hyperparameter tuning for all the machine learning methods was done using the GridSearchCV option in scikit-learn[38]. Hyperparameters for random forest such as number of descriptors, maximum depth of each tree were sampled and bootstrapping was permitted. The computational time for running the machine learning methods were on $O$(mins), which is astronomically lower than the alternative option of DEM simulations.

**Figures**

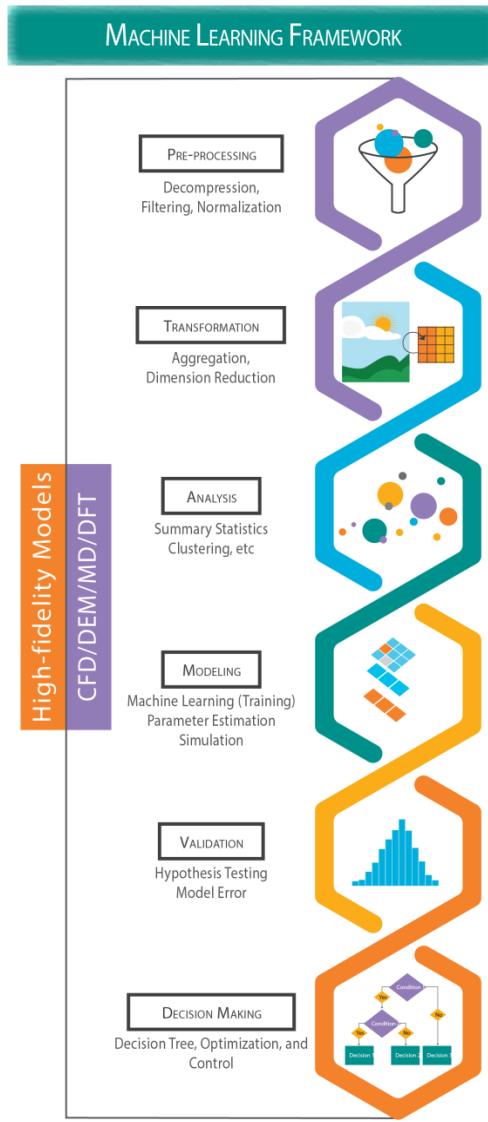

**Figure 1: Flowchart of steps involved in applying machine-learning to computationally expensive high-fidelity scientific models.** Availability to high-quality data is key to develop a good machine learning predictive model. Data transformation and feature engineering enables advanced data inspection, also contributing to better model performance.

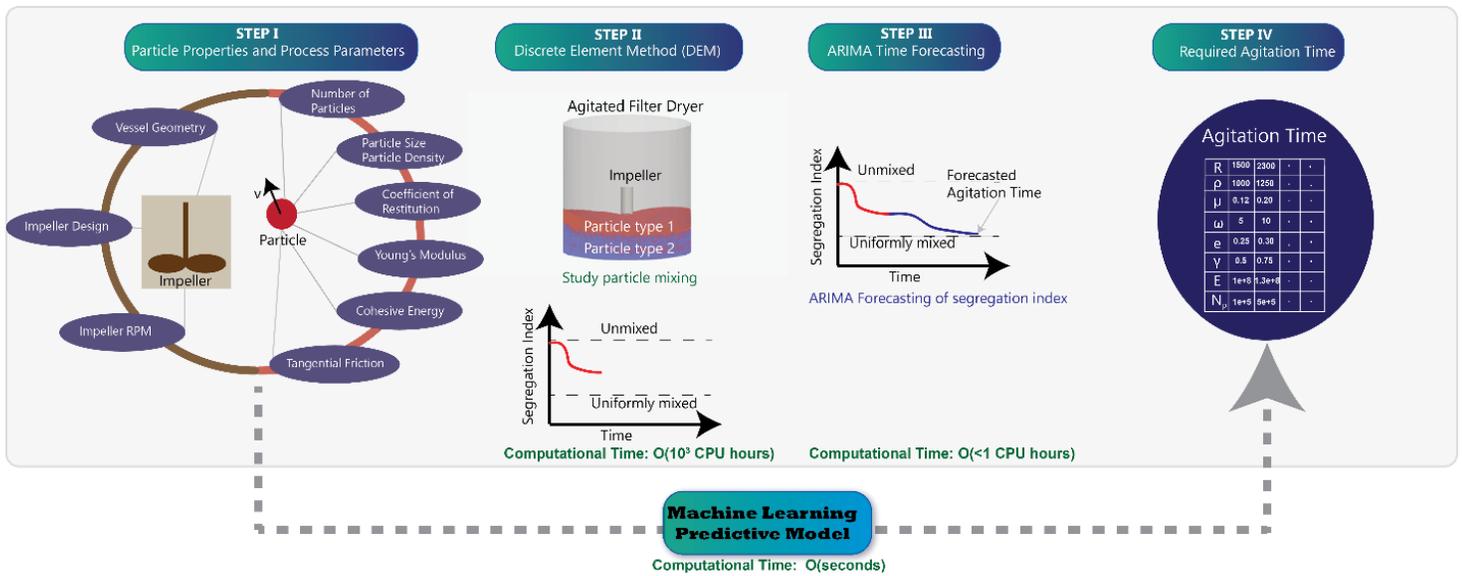

**Figure 2: Flowchart of the DAMPMix (DEM-ARIMA-Machine Learning Prediction of Mixing) approach to model solid particle mixing.** DEM simulations should be carried out for some initial time steps to provide the training data for the ARIMA model. ARIMA can then be implemented to forecast the mixing behavior and compute the required agitation time. Finally, a machine learning model can be built to predict the agitation time for any set of material properties and process parameters.

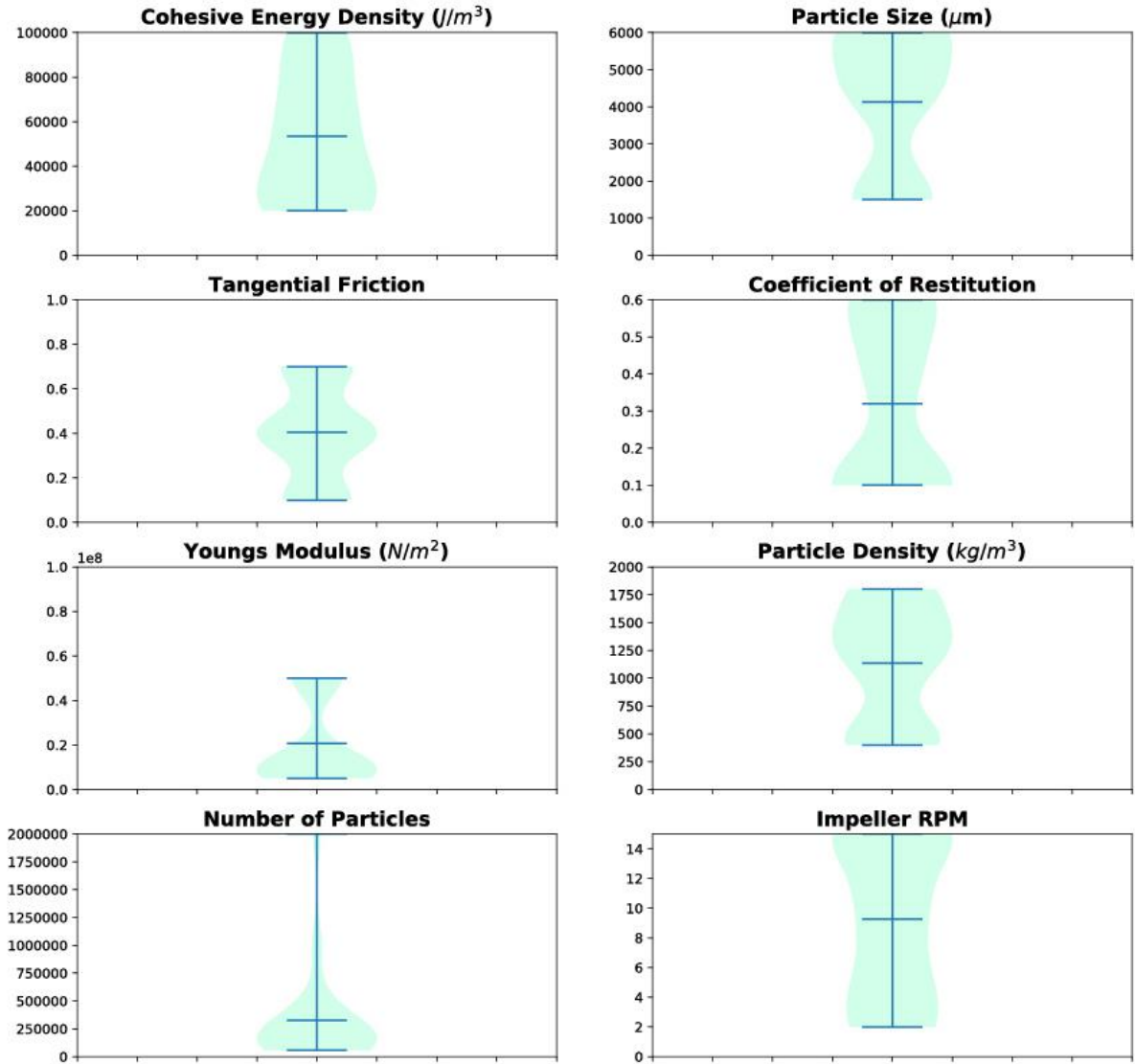

**Figure 3: Violin plot of the variation in material properties and process parameters, collectively known as Predictor variables.** In each plot, the second horizontal line (out of the three lines) shows the mean value of the individual material property, and the thickness shows the frequency of that particular value across all the simulations. It can be seen that there is good variability in the values of most of the properties except for number of particles, which can be attributed to the computational challenges of simulating larger number of particles.

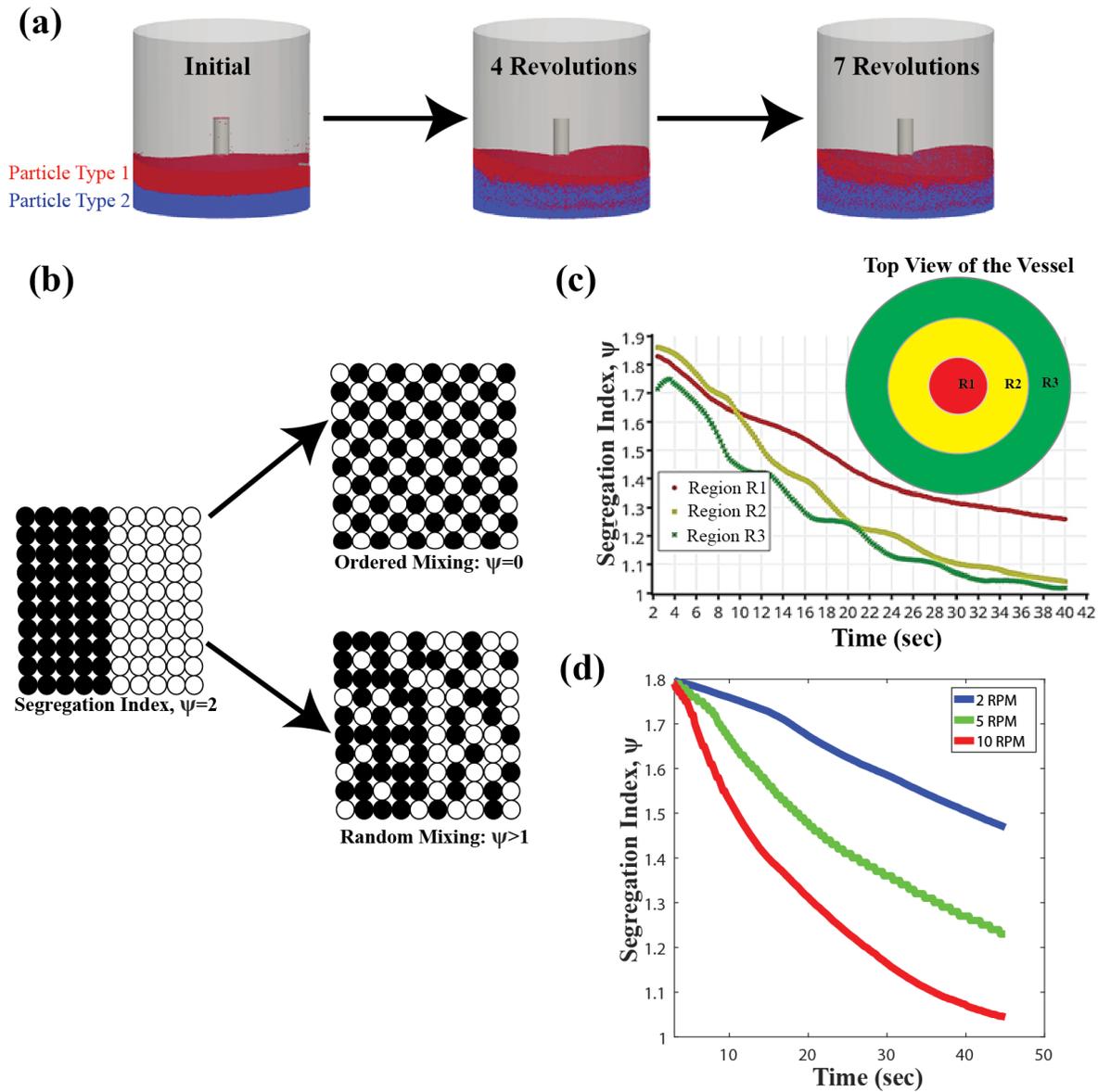

**Figure 4: (a) Extent of particle mixing with number of impeller rotations.** R = 3 mm, RPM = 15, E = 5 × 10⁷ N/m², $\gamma_{cohesion} = 1 \times 10^5$ J/m³, $\mu_f = 0.1$, ρ = 1100 kg/m³, e = 0.6, h = 20.33 cm. Particles are labeled by two types to examine their mixing behavior, even though their properties are same, (b) Different particle arrangements and the corresponding segregation index[23] (c) Particle mixing is faster in regions further from the center of the impeller. Region R1, R2 and R3

span the radial direction of the bed with R1 being the closest to the center of the impeller and R3 being closes to the dryer wall, and (d) Particle mixing is a function of the number of impeller revolutions. Longer simulations are required for slower RPMs.

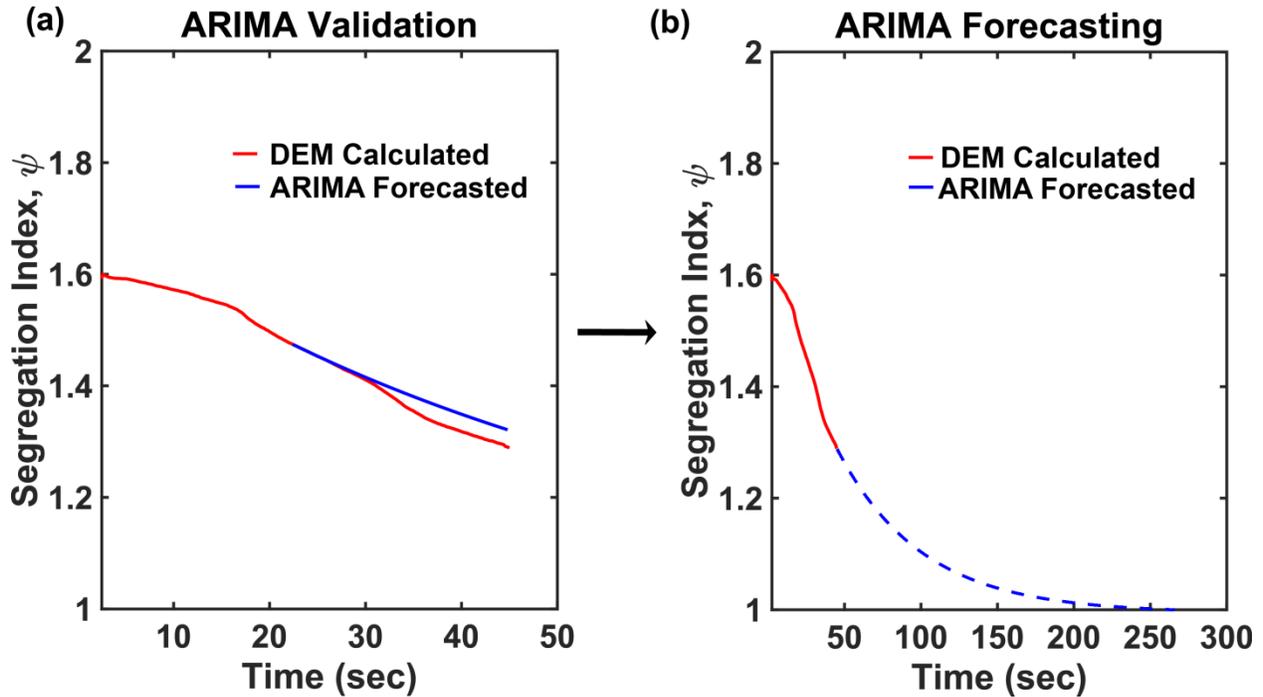

**Figure 5**: **Validation of ARIMA time-series forecasting**. (a) ARIMA was verified against the predictions due to DEM simulations for impeller speed of 2 RPM, and (b) ARIMA was used to forecast ψ till the bed was uniformly mixed.

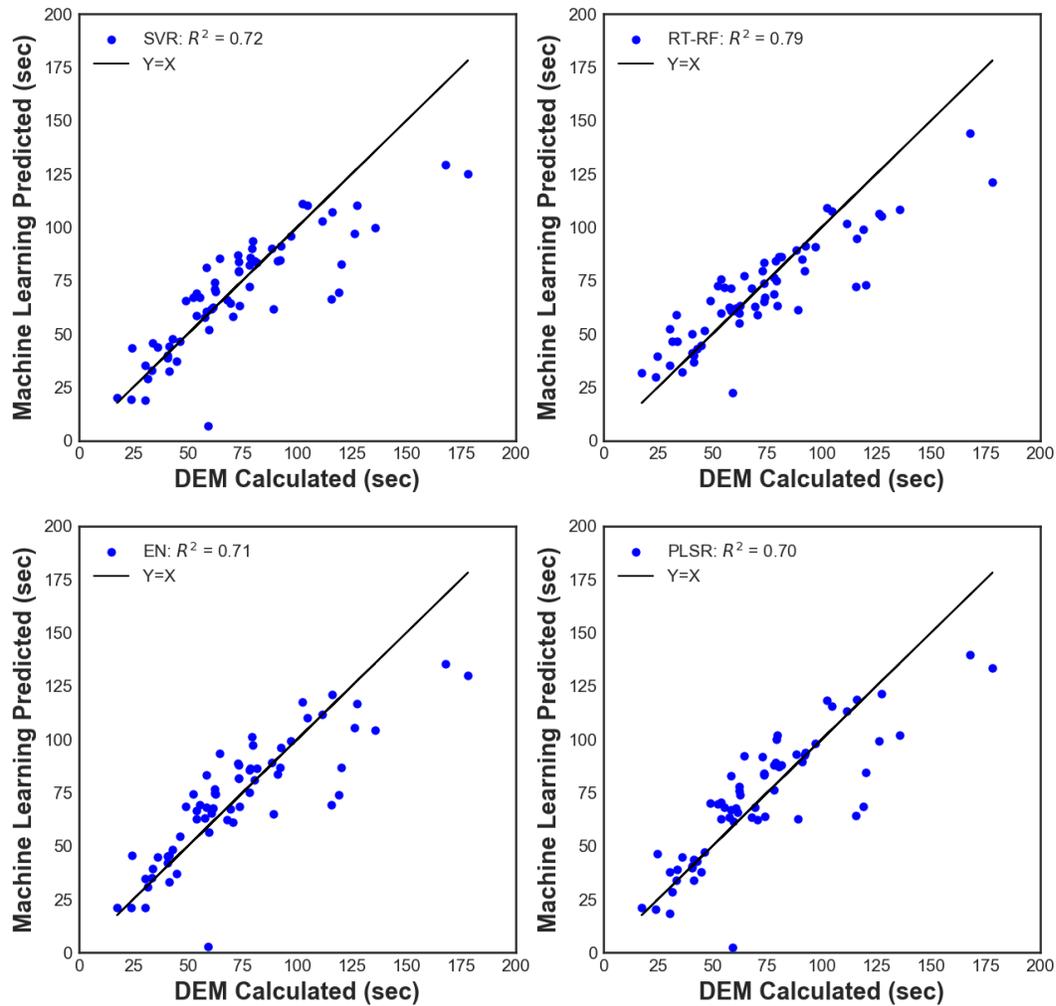

**Figure 6: ML prediction of agitation time compared to DEM-ARIMA simulations.** Leave-one-out cross-validation was performed to evaluate the methods. RT-RF performed the best amongst all the methods with an R$^2$ of 0.79.

**Author Contributions**
**Prashant Kumar** (lead author) worked on the machine learning and time-series analysis of the study. He also ran and analyzed the DEM simulations. Prashant worked on drafting the manuscript and generated all the figures.

**Kushal Sinha** (corresponding author) conceived, designed and oversaw the entire work, ran DEM simulations, and worked on drafting the manuscript.

**Nandkishor Nere** and **Laurie Mlinar** helped in designing and guiding the work by providing practical insights from their operational experiences which guided the DEM simulation work. They also reviewed the manuscript, and substantially revised it.

**Yujin shin** and **Raimundo Ho** helped in designing and guiding the work by providing insight in range of material properties and their behavior in pharmaceutical operations which guided DEM simulations and subsequent ML work. They also reviewed the manuscript and substantially revised it.

**Ahmad Sheikh** reviewed the manuscript and helped in its restructuring and revisions.

**Additional information**

All authors are AbbVie employees and may own AbbVie stock. AbbVie sponsored and funded the study; contributed to the design; participated in the collection, analysis, and interpretation of data, and in writing, reviewing, and approval of the final manuscript. The author(s) declare no competing interests.